\documentstyle[a4,german]{article}

\title{A Tool to Recover Scalar Time-Delay Systems from Experimental Time Series}
\author{M. J. B"unner, M. Popp, Th. Meyer, A. Kittel, J. Parisi
        \thanks{published in Phys. Rev. E {\bf 54} (1996) R3082.}\\ 
       {\em Physical Institute, University of Bayreuth,
            D-95440 Bayreuth, Germany}}
\date{April, 24th, 1996}

\begin{document}


\large
\maketitle


\begin{abstract}
We propose a method that is able to analyze chaotic time series, gained from experimental 
data. The method 
allows to identify scalar time-delay systems. If the dynamics of the system 
under investigation is governed by a scalar time-delay differential equation of the form 
$\frac{dy(t)}{dt} = h(y(t),y(t-\tau_0))$, the delay time $\tau_0$ and the function $h$ can 
be recovered. There are no restrictions to the dimensionality of the chaotic
attractor. The method turns out to be  insensitive to noise. We successfully 
apply the method to various time series taken from a computer experiment 
and two different electronic oscillators. 


P.A.C.S.: 05.45.+b
\end{abstract}



Time series analysis of chaotic systems has gained much interest in 
recent years.  Especially, 
embedding of time series in a reconstructed phase space with the 
help of time-delayed 
coordinates  was widely used to estimate fractal dimensions of 
chaotic attractors \cite{takens,grassberger}
and Lyapunov exponents \cite{wolf}. It is the advantage of 
embedding techniques that the 
time series of only one variable has to be analyzed, even if the 
investigated system is 
multi-dimensional. Furthermore,  it can be applied, in 
principle,  to any dynamical system. 
Unfortunately, the embedding techniques only yield information, if 
the dimensionality of the 
chaotic attractor under investigation is low. Another drawback is 
the fact that it does not 
give any information about the structure of the dynamical system, in the 
sense, that one is able to 
identify the underlying instabilities. In the following, we propose a method which is 
taylor-suited to identify scalar systems with a time-delay induced instability. We will show 
that the 
differential equation can be recovered from the time series, if the 
investigated dynamics obeys a scalar time-delay differential 
equation. There are no restrictions to the 
dimensionality of the chaotic attractor. Additionally, the method 
has the advantage to be insensitive to noise. 



We consider the time evolution of scalar time-delay differential 
equations 
\begin{equation}
\label{tdde}
\dot{y}(t) = h(y(t),y(t-\tau_0)), 
\end{equation}
with the initial condition
\begin{equation}
y(t)=y_0(t), \hspace{1.0cm} -\tau_0<t<0. \nonumber
\end{equation}
The dynamics is supposed to be bounded in the counter-domain 
$\cal{D}$, $y(t) \in \cal{D}, \forall$ $t$. 
In equation (\ref{tdde}), the time derivative of $y(t)$  does not only depend on 
the state of system at the time $t$, but there also exist nonlocal 
correlations in time, because the function $h$  additionally 
depends on the time-delayed value $y(t-\tau_0)$. These 
nonlocal correlations in time enable scalar time-delay systems to 
exhibit a complex 
time evolution. The 
number of positive Lyapunov exponents increases with the delay time 
$\tau_0$ \cite{farmer}. Scalar time-delay systems, therefore, constitute a major class of 
dynamical 
systems which exhibit 
hyperchaos \cite{roessler}. In general, though, the nonlocal correlations in 
time are not at all obvious from the 
time series. 
A state of the system (\ref{tdde}) is uniquely defined by a function on
an interval of length $\tau_0$.
Therefore, the phase space of scalar time-delay systems must be 
considered as 
infinite dimensional.  
The trajectory in the infinite dimensional phase space $\vec{y}(t)=\{y(t'), 
t-\tau_0<t'<t\}$ is easily obtained from the time series. The scalar time series $y(t)$, 
therefore, encompasses the complete information about the trajectory $\vec{y}(t)$ in the 
infinite dimensional phase space.


The main idea of our analysis method is the following.  We project 
the trajectory   $\vec{y}(t)$ from the infinite dimensional phase space to a 
three-dimensional space 
which is spanned by the coordinates 
$(y_{\tau_0}=y(t-\tau_0),y=y(t),\dot{y}=\dot{y}(t))$. In 
the  $(y_{\tau_0},y,\dot{y})$-space the differential 
equation (\ref{tdde}) determines a two-dimensional surface $h$. The projected
trajectory $\vec{y}_{\tau_0}(t)=(y(t-\tau_0),y(t),\dot{y}(t))$, 
therefore, 
is confined to the surface $h$ and is not able to explore other 
directions of the   
$(y_{\tau_0},y,\dot{y})$-space. 
From this, we conjecture, that the fractal dimension of the 
projected attractor has to be between one and two. Furthermore, it follows that any 
intersection of the chaotic 
attractor with a surface 
$k(y_{\tau_0},y,\dot{y})=0$ yields a curve. More precisely spoken, if 
one transforms 
the projected trajectory $\vec{y}_{\tau_0}(t)$ to a series of points  
$\vec{y}_{\tau_0}^i=(y^i_{\tau_0},y^i,\dot{y}^i)$ that fulfill the condition 
$k(y^i_{\tau_0},y^i,\dot{y}^i)=0$, the 
series of points  
$(y^i_{\tau_0},y^i,\dot{y}^i)$ contract to a curve and  its 
dimension has to be less than or 
equal to one. 
In general, it cannot be expected that one is able to project a 
chaotic attractor of arbitrary  dimension to a 
three-dimensional space, in the way that its projection is  embedded in a 
two-dimensional surface. We, nevertheless, demonstrate that this is always possible 
for chaotic attractors of scalar time-delay systems (\ref{tdde}).
 


In the following, we will show that such finding can be used to 
reveal 
nonlocal correlations in time from the time series. 
If the dynamics is of the scalar time-delay type (\ref{tdde}), the 
appropriate delay time 
$\tau_0$ 
and the function $h(y,y_{\tau_0})$ can be recovered . The 
trajectory in the infinite dimensional phase space $\vec{y}(t)$ 
is projected to several three-dimensional  
$(y_{\tau},y,\dot{y})$-spaces upon variation of $\tau$. The appropriate value  $\tau=\tau_0$ 
is just the one for which the projected trajectory $\vec{y}_{\tau}$ lies on a surface, 
representing  a 
fingerprint of the time-delay 
induced instability. Projecting the trajectory $\vec{y}$ 
to the $(y_{\tau_0},y,\dot{y})$-space, the projected trajectory yields 
the surface $h(y,y_{\tau_0})$ 
in the 
counter-domain $\cal{D} \times \cal{D}$. With a fit 
procedure the yet unknown function $h(y,y_{\tau_0})$ can be 
determined in $\cal{D} \times 
\cal{D}$. Therefore, the complete scalar time-delay differential 
equation has been recovered 
from the 
time series. 
In some cases, it is more convenient to intersect the trajectory $\vec{y}_{\tau}$ 
in the $(y_{\tau},y,\dot{y})$-space with 
a surface $k(y_{\tau},y,\dot{y})=0$ which yields a series of points 
$\vec{y}_{\tau}^i=(y^i_{\tau},y^i,\dot{y}^i)$.
For $\tau=\tau_0$, the points come to lie 
on a curve and the fractal 
dimension of the point set has to be less than or equal to one.  


The analysis method is also applicable if noise is added to the time series. The only
effect of additional noise is that the projected time series in the 
$(y_{\tau_0},y,\dot{y})$-space is not perfectly enclosed in a two-dimensional surface, but
the surface is somewhat blurred up. If the analysis is done with an intersected trajectory,
the alignment of the noisy data is not perfect. 
The arguments presented above do not require the dynamics to be
settled on its chaotic attractor. Therefore, it is also possible to analyze 
transient chaotic dynamics. Recently, the coexistence
of attractors of time-delay systems has been pointed out \cite{losson}.
The only requirement for the analysis method is that the trajectory obeys the 
time-evolution equation (\ref{tdde}) which holds for all 
coexisting attractors in a scalar time-delay system. Therefore, the method is
applicable, no matter in which attractor the dynamics has
decided to settle. The analysis requires only short time series, which makes
it well-suited to be applied on experimental situations.
We successfully apply the method to time series gained from a 
computer experiment and from two different  electronic oscillators. We show the robustness
of the method to additional noise by analyzing noisy time series.


We numerically calculated  the time series of the scalar 
time-delay differential equation
\begin{eqnarray}
\label{tdde2}
\dot{y}(t) & = & f(y_{\tau_0})- g(y),\\
f(y_{\tau_0}) & = & \frac{2.7y_{\tau_0}}{1+y_{\tau_0}^{10}} +c_0 \nonumber \\
g(y) & = &  -0.567y + 18.17y^2 -38.35y^3+28.56y^4-6.8y^5 -c_0 \nonumber
\end{eqnarray}
with the initial conditon
\begin{equation}
y(t) = y_0(t),\hspace{1.0cm} -\tau_0<t<0,\nonumber 
\end{equation}
which is of the form (\ref{tdde}) with $h(y_{\tau_0},y)=f(y_{\tau_0})-g(y)$. 
The function $g$ has been chosen to be non-invertible in the counter-domain $\cal{D}$. The
definition of the functions $f$ and $g$ is ambiguous in the sense that adding
a constant $c_0$ to $f$ can always be cancelled by subtracting $c_0$ from $g$ without 
changing 
$h$ and, therefore, leaving the dynamics of equation 
(\ref{tdde2}) 
unchanged.  The 
control parameter is the delay time $\tau_0$.
Equation (\ref{tdde2}) is somewhat similar to the Mackey-Glass equation 
\cite{mkg}, except for the 
function $g$, which is linear in the Mackey-Glass system. Part of 
the time series is shown 
in Fig. 1. We used $500,000$ data points with a time step of $0.01$ for the analysis. The 
dimension of the chaotic attractor was 
estimated with the help of the 
Grassberger-Procaccia algorithm \cite{grassberger} to be clearly larger than $5$ . 
To recover the delay time $\tau_0$ and the 
functions $f$ and $g$ from the time series, we applied the analysis 
method outlined above. We projected the trajectory $\vec{y}(t)$ from the 
infinite dimensional phase space  to several
$(y_{\tau},y,\dot{y})$-spaces under variation of $\tau$ and intersected the projected 
trajectory $\vec{y}_{\tau}$
with the $(y=1.1)$-plane, which is repeatedly traversed by the trajectory, as can be seen in 
Fig. 1. 
The results are the times $t^i$ where the trajectory traverses the 
$(y=1.1)$-plane and the intersection points
$\vec{y}_{\tau}^i=(y^i_{\tau},1.1,\dot{y}^ i)$.  For $\tau$ 
being the appropriate value $\tau_0$, the point set 
$\vec{y}_{\tau_0}^i$ is correlated via 
equation (\ref{tdde2}) 
\begin{equation}
\label{f}
\dot{y}^i =  f(y^i_{\tau_0})- g(1.1)
\end{equation}
and, therefore, must have a fractal dimension less than or equal to 
one. Then, we ordered 
the  $(y^i_{\tau},\dot{y}^ i)$-points with respect to the values of $y^i_{\tau}$. 
A simple measure for 
the alignment of the points is the length $L$ of a polygon 
line connecting all ordered 
points  $(y^i_{\tau},\dot{y}^ i)$.
The length $L$ as a function of $\tau$ is shown in Fig. 2. For 
$\tau=0$, $L(\tau)$ is minimal, 
because the points $(y^i_\tau,\dot{y}^ i)$ are ordered along the 
diagonal in the 
$(y^i_\tau,\dot{y}^ i)$-plane.  $L(\tau)$ increases with $\tau$ and 
eventually reaches a 
plateau, where the points $(y^i_\tau,\dot{y}^ i)$ are maximally uncorrelated. 
This is due to short-time correlations of the signal. Eventually, $L(\tau)$ decreases 
again and
shows a  dip for $\tau$ reaching the appropriate value 
$\tau_0$. 
A further decrease of $L(\tau)$ is observed for $\tau=2\tau_0$.
In Fig. 3(a)-(c), we show the projections $\vec{y}_{\tau}(t)$ of the
trajectory $\vec{y}(t)$ from the infinite dimensional phase space to different  
$(y_{\tau},y,\dot{y})$-spaces under variation of $\tau$.  Clearly, for $\tau$ 
approaching the appropriate value 
$\tau_0$, the appearance of the projected trajectory changes. 
In Fig. 3(c), the projected trajectory is embedded in  a surface which is determined by the 
function $h$.
In Fig. 3(d)-(f) we show the 
point set $(y^i_\tau,\dot{y}^ i)$ resulting from the intersection of the 
projected trajectory $\vec{y}_{\tau}$ with the  
$(y=1.1)$-plane. The point set is projected to the 
$(y_\tau,\dot{y})$-plane. According to equation (\ref{f}), the 
points are aligned along the function $f$ for $\tau=\tau_0$. 
With the appropriate value $\tau_0$, we are in the position to 
recover the functions $f$ and 
$g$ from the time series. 
The functions $f$ and $g$ are ambiguous with respect to the addition of a constant term 
$c_0$, 
as has 
been outlined above. Therefore, one is free to remove the ambiguity by invoking an additional 
condition which we choose to be
\begin{equation}
\label{norm}
g(1.1)=0. 
\end{equation}
Then, equation (\ref{f}) reads
\begin{equation}
\label{f_r}
\dot{y}^i =  f(y^i_{\tau_0}).
\end{equation}
Therefore, function $f$ is recovered by analyzing the intersection points 
$\vec{y}_{\tau_0}^i$ in  the 
$(\dot{y},y_{\tau_0})$-plane. To recover the function $g$, we intersected the time series 
with the $(y_{\tau_0}=1.1)$-plane. The resulting point set 
$\vec{y}_{\tau_0}^j=(\dot{y}^j,y^j)$ is correlated via
\begin{equation}
\label{g_r}
\dot{y}^j =  f(1.1)-g(y^j).
\end{equation}
The value $f(1.1)$ has been taken from the time series using equation (\ref{f_r}).
In Fig. 4(a)-(b), we compare the functions $f$ and $g$, as they have been defined in equation 
(\ref{tdde2}) with the recovery of the functions  $f$ and $g$ from the time series. We 
emphasize that no 
fit parameter is involved. 


We checked the robustness of the method to additional noise by analyzing noisy time
series, which had been produced by adding gaussian noise to the time series of equation 
(\ref{tdde2}). We analyzed two noisy time series with a signal-to-noise ratio (SNR) of $10$ 
and 
$100$. In both cases, the additional noise was partially removed with a 
nearest-neighbor filter (for SNR = 100, average over six neighbors; for SNR = 10, average 
over 
twenty neighbors). After that, the noisy time series were analyzed in the same way as
has been described above. The inset of Fig. 2 shows the result of the
analysis. The length $L$ of the polygon line exhibits a local minimum for $\tau=\tau_0$. In 
the 
case
of the time series with a SNR of $10$, the local minimum is again sharp, 
but 
somewhat less pronounced. We conjecture that the method is robust
with respect to additional noise and, therefore, well suited for the
analysis of experimental data. 


Finally, we successfully applied the method to experimental time 
series gained from two different types of electronic oscillators. 
The first one is the Shinriki oscillator \cite{shinriki,reisner}.
The 
dynamics of the second oscillator \cite{pyragas} is time-delay 
induced and mimics the dynamics 
of the Mackey-Glass equation. In both cases, we intersected the 
trajectory with the $(\dot{y}=0)$-plane. The resulting point set 
was represented in a $(y_\tau,y)$-space with 
different values of $\tau$. Then, we ordered the points with respect to 
$y_\tau$ and the length $L$ of 
a polygon line connecting all ordered points  $(y^i_\tau,y^i)$ was 
measured. The results are
presented in Fig. 5 (a) and Fig. 5 (b). In both cases, $L(\tau)$ has a local
minimum for small values of $\tau$ as a result 
of short-range correlations in time. $L(\tau)$ increases in
time and reaches a plateau. For the Shinriki oscillator,  no 
further decrease of $L(\tau)$ is 
observed for increasing $\tau$ (Fig. 5(a)). Such finding clearly shows 
that the dynamics of 
the Shinriki oscillator is not time-delay induced. Analyzing the 
Mackey-Glass oscillator (Fig. 
5(b)), one finds sharp dips in $L(\tau)$ for $\tau=\tau_0$ and  
$\tau=2\tau_0$. This is a 
direct evidence for correlations in time, which are induced by the time delay (for details see 
\cite{physletta96}). 
Obviously, the 
method is able to identify 
nonlocal correlations in time from the time series.
Eventually, the 
nonlinear characteristics of the electronic oscillator is compared to its
recovery from the time series (Fig. 5 (c)).   


In conclusion, we have presented a method capable to reveal nonlocal 
correlations in time of scalar systems 
by analyzing the time series. If the dynamics of the investigated 
system is governed by a 
scalar time-delay differential equation, we are able to recover the 
scalar time-delay differential 
equation. There are no constraints on the 
dimensionality of the attractor. Since scalar time-delay systems are able to exhibit
high-dimensional chaos, our method might pave the road
to inspect high-dimensional chaotic systems, where conventional 
time-series analysis techniques already fail. Furthermore, the motion is not required to 
be settled on its 
attractor. The method is insensitive 
against 
additional noise. We have successfully 
applied the method to time series gained from a computer experiment 
and to experimental data 
gained from two different types of  electronic oscillators.  


While, in general, the verification of dynamical models is a highly complicated
task, we have shown that the identification of scalar time-delay systems can be
accomplished easily and, thus,  allows a detailed comparison of the model equation
with experimental time series.
In 
several disciplines, e.g., hydrodynamics \cite{villermaux95} , chemistry 
\cite{khrustova95}, laser physics \cite{ikeda87}, and physiology \cite{mkg,longtin90}, 
time-delay effects have been proposed to induce dynamical 
instabilities. With the help of our method, there is a good 
chance to verify these models by analyzing the 
experimental time series. If the dynamics is indeed governed by a time delay, the
delay time and the time-evolution equation can be determined.  
Current and future research activities of the authors concentrate on extending the time-series 
analysis method
to non-scalar time-delay systems as well as to time-delay systems with multiple delay times.



We thankfully acknowledge valuable discussions with J. Peinke and K. Pyragas and financial 
support of the 
Deutsche Forschungsgemeinschaft.





\section*{Figure captions}
\begin{itemize}
\item[{\bf Fig. 1:}]
	Time series of the scalar time-delay system (\ref{tdde2}) obtained from a 
computer experiment ($\tau_0=40.00$). 


\item[{\bf Fig. 2:}]
	Length $L$ of the polygon line connecting all ordered 
points of the projected 
point set $(y^i_\tau,\dot{y}^ i)$ versus $\tau$. $L$ has been normalized so that a maximally 
uncorrelated 
point set has the value
$L=1.0$. The inset shows a close-up of the $\tau$-axis around the local minimum at 
$\tau=\tau_0=40.00$. Additionally, $L(\tau)$-curves gained from the analysis of noisy time 
series 
are shown (no additional noise -- straight line, signal-to-noise ratio of $100$ -- open 
circles, and
signal-to-noise ratio of $10$ -- squares).


\item[{\bf Fig. 3:}]
	(a)-(c): Trajectory $\vec{y}_{\tau}(t)$ which has been projected from the infinite 
dimensional phase space to the 
$(y_{\tau},y,\dot{y})$-space under variation of $\tau$. (a) $\tau=20.00$. (b) $\tau=39.60$. 
(c) 
$\tau=\tau_0=40.00$. (d)-(f): Projected 
point set $\vec{y}_{\tau}^i=(y^i_\tau,\dot{y}^ i)$ resulting from the intersection of the 
projected trajectory $\vec{y}_{\tau}(t)$ with the 
$(y=1.1)$- plane under variation of $\tau$. (d) $\tau=20.00$. (e) $\tau=39.60$. (f) 
$\tau=\tau_0=40.00$.


\item[{\bf Fig. 4:}]
	(a) Comparison of the function $f$ (line) of equation 
(\ref{tdde2}) with its recovery 
 from the time series (points). (b) Comparison of the 
function $g$ (line) of equation 
(\ref{tdde2}) with its recovery   from the time series (points).


\item[{\bf Fig. 5:}]
	Length $L$ of the polygon line connecting all ordered 
points of the projected 
point set $\vec{y}_{\tau}^i=(y^i_\tau,y^ i)$ versus $\tau$ for (a) the Shinriki and 
(b) the Mackey-Glass 
oscillator. $L(\tau)$ has been normalized so that it has the value $L=1$ for an 
uncorrelated point set.
(c) Comparison of the nonlinear characteristics of the Mackey-Glass oscillator, which is
the function $f(y_{\tau_0})$ of an ansatz of the form $h(y,y_{\tau_0}) = f(y_{\tau_0}) + g(y)$, 
measured directly on the oscillator (line) with its recovery from the time series (dots).
\end{itemize}



\begin{thebibliography}{XXX}
\bibitem{takens}F. Takens, Lect. Notes Math. {\bf 898} (1981) 366.
\bibitem{grassberger}P. Grassberger, I. Procaccia, Physica D {\bf 
9} (1983) 189.
\bibitem{wolf} A. Wolf, J. B. Swift, H. L. Swinney, J. Vastano, 
Physica D {\bf 16} (1985) 285.
\bibitem{farmer} J. D. Farmer, Physica D {\bf 4} (1982) 366.
\bibitem{roessler}O. E. R\"ossler, Z. Naturforsch. {\bf 38a} (1983) 
788.
\bibitem{losson} J. Losson, M. C. Mackey, A. Longtin, Chaos (AIP), {\bf 3} 
(1993), 167.
\bibitem{mkg}M. C. Mackey, L. Glass, Science {\bf 197} (1977) 287.
\bibitem{shinriki} M. Shinriki, M. Yamamoto, S. Movi, Proc. IEEE 
{\bf 69} (1981) 394.
\bibitem{reisner} B. Reisner, A. Kittel, S. L\"uck, J. Peinke, J. Parisi, Z. Naturforsch.{\bf 
50a} (1995) 105.
\bibitem{pyragas} A. Namajunas, K. Pyragas, A. Tamasevicius, Phys. 
Lett. A {\bf 201} (1995) 42.
\bibitem{physletta96} M. J. B\"unner, M. Popp, Th. Meyer, A. Kittel, U. Rau, J. Parisi,
Phys. Lett. A {\bf 211} (1996) 345.
\bibitem{villermaux95} E. Villermaux, Phys. Rev. Lett. {\bf 75} (1995) 4618. 
\bibitem{khrustova95} N. Khrustova, G. Veser, A. Mikhailov, Phys. Rev. Lett. {\bf 75}
 (1995) 3564.
\bibitem{ikeda87} K. Ikeda, K. Matsumoto, Physica D {\bf29} (1987) 223.
\bibitem{longtin90} A. Longtin, J. G. Milton, J. E. Bos, M. C. Mackey, Phys. Rev. A {\bf 41}
(1990) 6992.
\end{thebibliography}
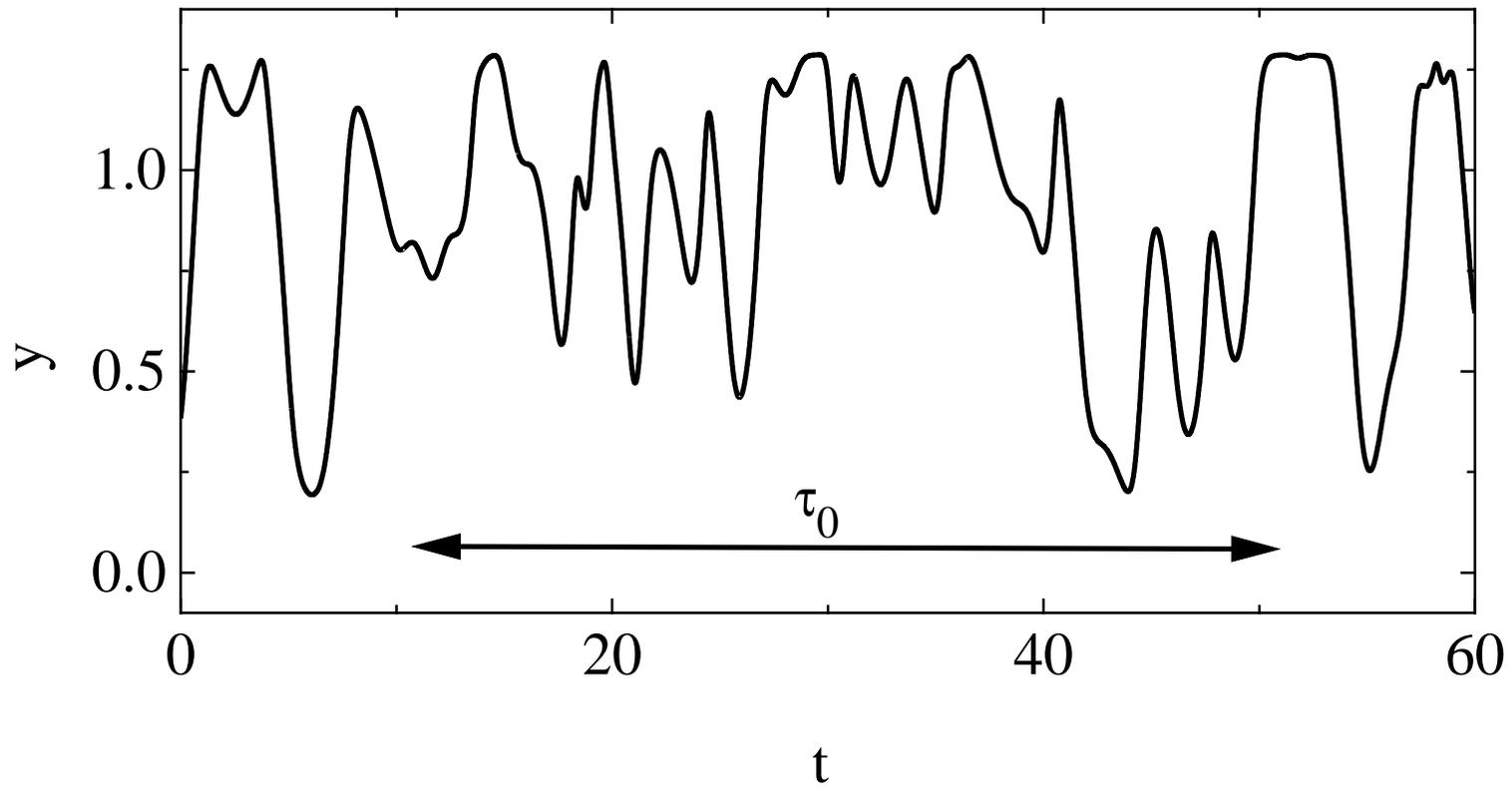
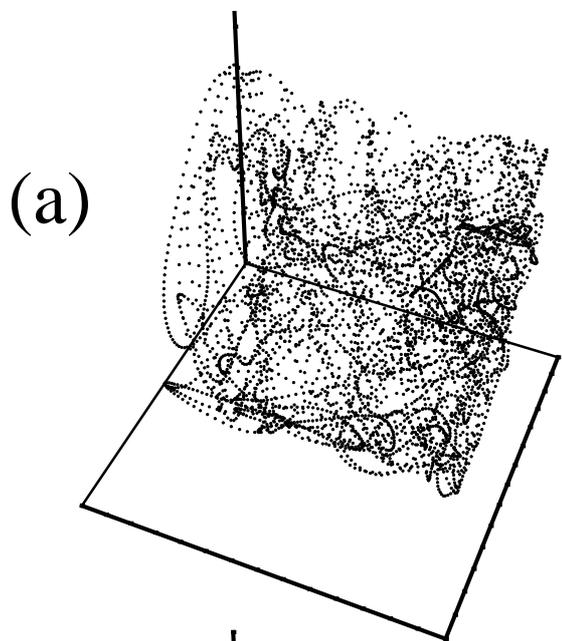
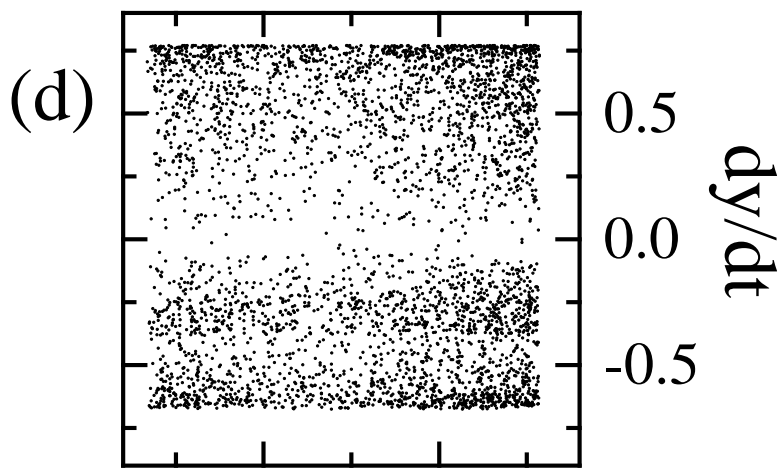
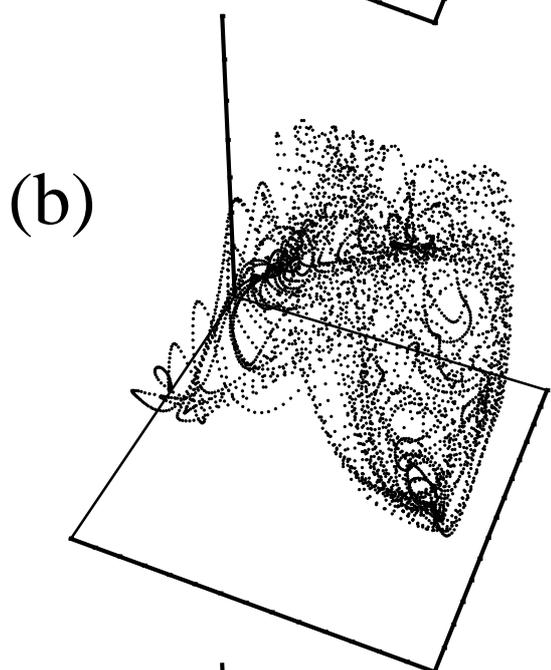
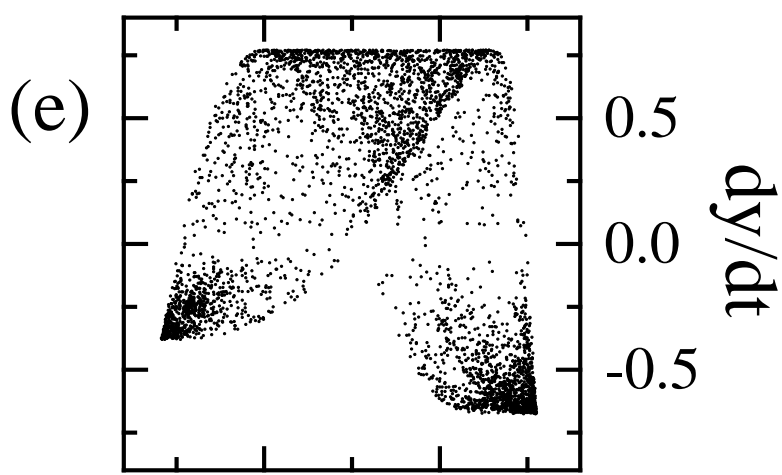
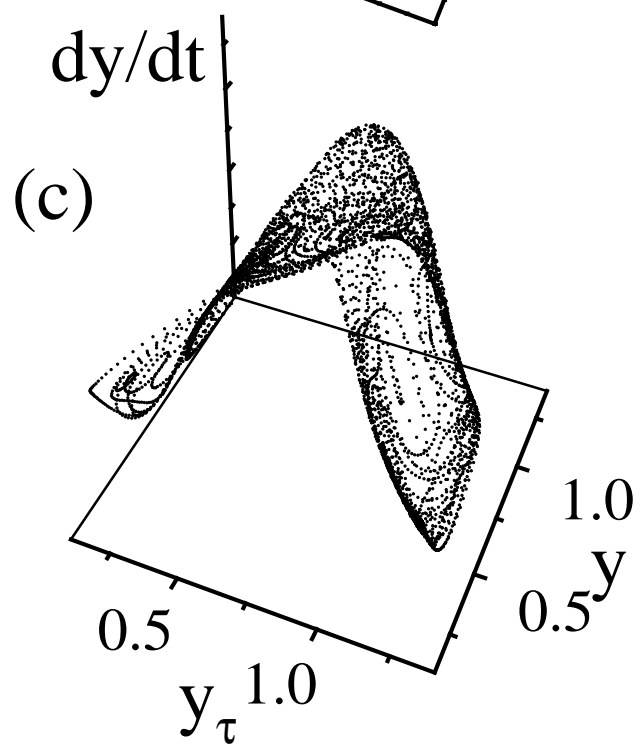
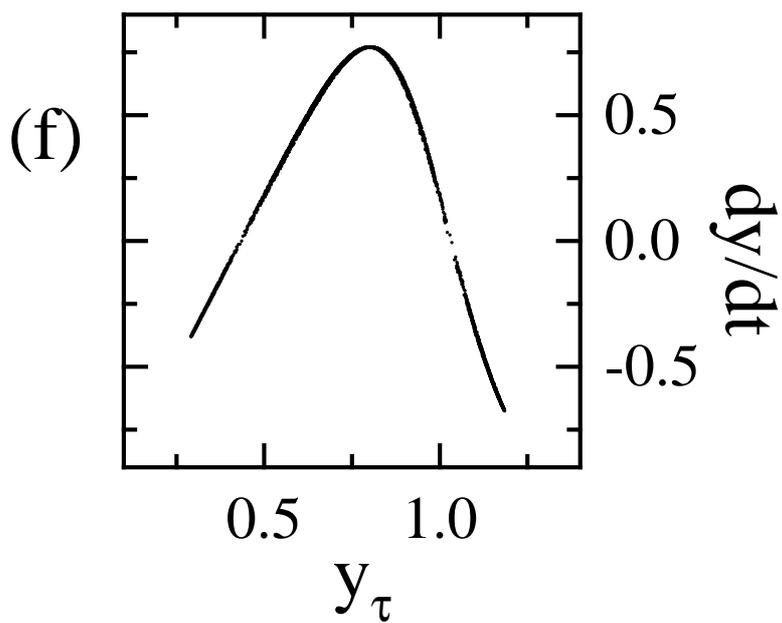
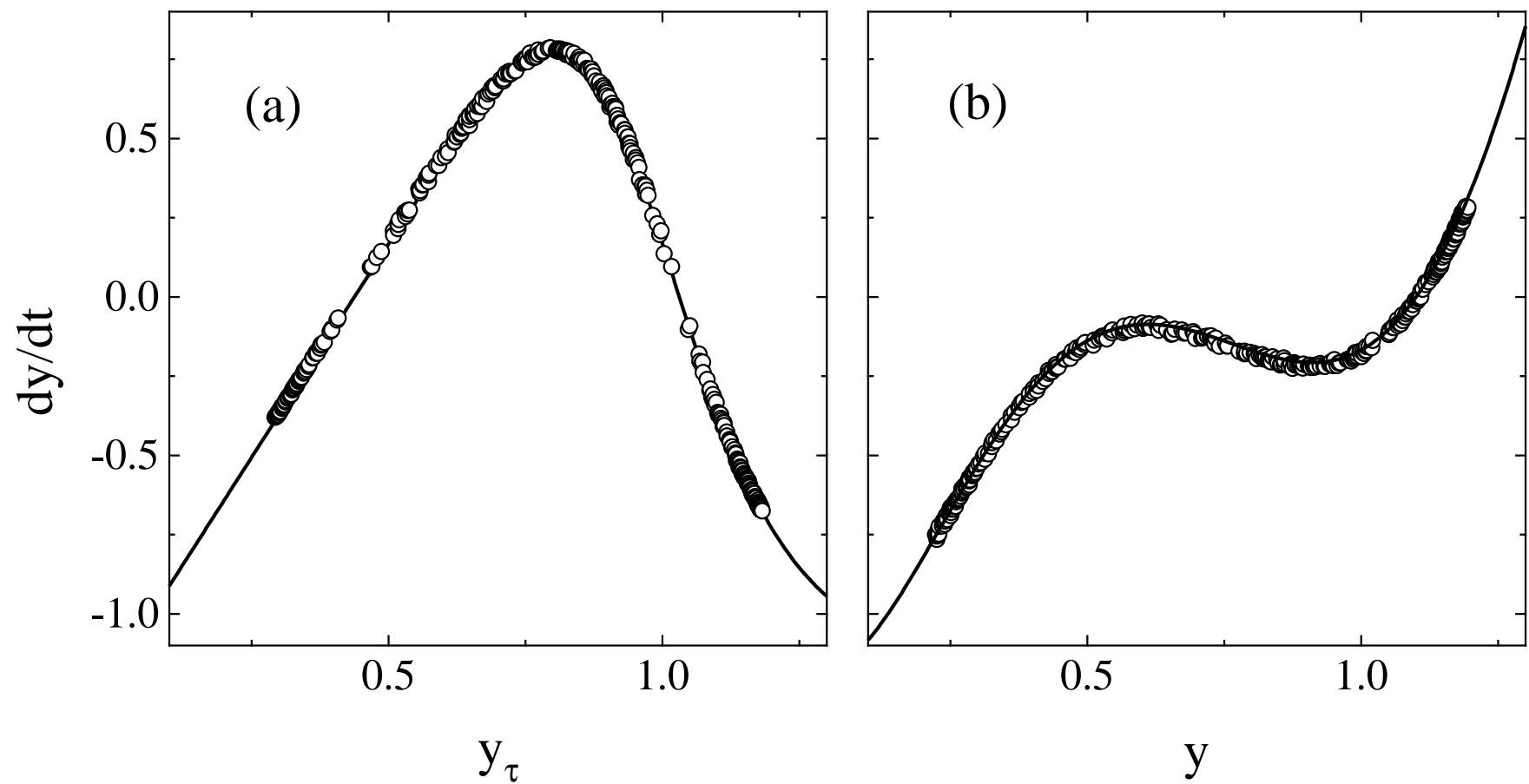
\end{document}